\documentclass[aps,preprintnumbers,superscriptaddress,showpacs]{revtex4}
\usepackage{epsfig}
\usepackage{psfrag}
\usepackage{amsfonts}
\usepackage{graphicx}
\usepackage{dcolumn}
\usepackage{bm}

\begin{document}

\title{Imaginary potential of moving quarkonia in a D-instanton background}

\author{Zi-qiang Zhang}
\email{zhangzq@cug.edu.cn} \affiliation{School of mathematics and
physics, China University of Geosciences(Wuhan), Wuhan 430074,
China}

\author{De-fu Hou}
\email{houdf@mail.ccnu.edu.cn} \affiliation{Key Laboratory of
Quark and Lepton Physics (MOE), Central China Normal University,
Wuhan 430079,China}

\author{Gang Chen}
\email{chengang1@cug.edu.cn} \affiliation{School of mathematics
and physics, China University of Geosciences(Wuhan), Wuhan 430074,
China}

\begin{abstract}
The imaginary part of the inter-quarks potential of moving heavy
quarkonia is investigated in a dual supergravity of the AdS
background deformed by dilaton, which induces the gauge field
condensate in the dual gauge theory. We analyze the quark
anti-quark pair moving transverse and parallel to the plasma wind,
respectively. It is shown that for both cases the presence of
D-instanton density tends to increase the inter-distance and
decrease the imaginary potential, reverse to the effect of the
pair's velocity. Moreover, it is found that the D-instanton
density has stronger effects for the parallel case rather than
transverse.

\end{abstract}

\maketitle
\section{Introduction}
The experimental program at RHIC and LHC have produced a new state
of matter so-called quark gluon plasma (QGP) \cite{JA,KA,EV}. One
important experimental signal for QGP formation is melting of
quarkonia. It was suggested earlier that the main mechanism
responsible for this suppression is color screening \cite{TMA}.
But recently some authors argued that the imaginary part of the
potential, ImV$_{QQ}$, can yield this suppression as well
\cite{ML,AB,NB,MAE}. In addition, this quantity can be used to
estimate the thermal width of quarkonia. In the framework of
weakly coupled theories, ImV$_{QQ}$ has been studied in many
works, see e.g. \cite{NB1,AD,MM,VC}. As we know, at very high
temperatures (and/or densities), QCD may be treated
perturbatively, for instance, perturbative calculations taking
into account hard thermal loops are actually able to reproduce
lattice QCD results for the equation of state for temperatures
above T$\sim$300 MeV (see. e.g. \cite{NAJ}). However, at lower
temperatures, for example, around the QCD crossover region,
T$\sim$155 MeV, the perturbative calculation may not be fully
trusted, implying that nonperturbative methods are required in
this case. On the other hand, much experiment data indicates that
QGP is strongly coupled and behaves like a nearly idea fluid
\cite{EV,UW,SR}. Therefore, it would be interesting to study the
ImV$_{QQ}$ via the use of non-petubative methods, such as the
AdS/CFT correspondence
\cite{Maldacena:1997re,Gubser:1998bc,MadalcenaReview}.

AdS/CFT, the duality between a string theory in AdS space and a
conformal field theory in the physical space-time, has yielded
many important insights for studying different aspects of QGP
\cite{JCA}. In this approach, J. Noronha and A. Dumitru have
studied the imaginary potential and thermal width of quarkonia for
$\mathcal{N}=4$ SYM theory in their seminal work \cite{JN}.
Therein, the ImV$_{QQ}$ is related to the effect of thermal
fluctuations due to the interactions between the heavy quarks and
the medium. Later, this idea has been extended to various cases.
For example, the ImV$_{QQ}$ of static quarkonia is investigated in
\cite{JN1,KB}. The ImV$_{QQ}$ of moving quarkonia has been
analyzed in \cite{JN2,MAL}. The finite 't Hooft coupling
corrections on ImV$_{QQ}$ is discussed in \cite{KB1}. The
influence of chemical potential on ImV$_{QQ}$ has been addressed
in \cite{ZQ}. The study of ImV$_{QQ}$ in some AdS/QCD models, can
be found, for example, in \cite{NR,JS}. Also, there are other ways
to study ImV$_{QQ}$ from AdS/CFT, see \cite{JLA,THA}.

Now we would like to give such analysis in a D3/D-instanton
background, which corresponds to the Yang-Mills theory in the
deconfining, high-temperature phase. The background geometry
contains a nontrivial dilaton giving a nonzero gluon condensation
$q\propto <TrF^{\mu\nu}F_{\mu\nu}>$, where $q$ is also regarded as
the D-instanton density \cite{LHH}. It was argued \cite{BG,KG}
that the features of the D3/D-instanton geometry are similar to
QCD at finite temperature. Thus, one expects that the results
obtained from this theory should shed qualitative insights into
analogous questions in QCD. Due to this characteristic, many
quantities have been studied in the D-instanton background, such
as phase transitions \cite{BG}, flavor quark \cite{KG}, heavy
quark potential and jet quenching parameter \cite{ZQ1}.

In this paper, we will investigate the imaginary potential of
moving quarkonia in the D-instanton background. More specifically,
we would like to see how the D-density affects ImV$_{QQ}$ in this
case. Also, it would be interesting to compare the effects of the
D-density with those of the velocity. These are the main
motivations of the present work.

The outline of the paper is as follows. In the next section, we
briefly review the geometry of the D-instanton background at
finite temperature. In section 3, we study the imaginary potential
for the pair moving transverse and parallel to the plasma wind one
by one. The last part is devoted to conclusion and discussion.


\section{background geometry}
Let us begin with a briefly review of the D-instanton background.
The geometry is a finite temperature extension of D3/D-instanton
background given in \cite{KG}. The background has a five-form
field strength and an axion field which couples to D3 and
D-instanton, respectively. In Einstein frame the ten dimensional
super-gravity action is given by \cite{GW,AK}
\begin{equation}
S=\frac{1}{\kappa}\int
d^{10}x\sqrt{g}(R-\frac{1}{2}(\partial\Phi)^2+\frac{1}{2}e^{2\Phi}(\partial\chi)^2-\frac{1}{6}F^2_{(5)}),\label{action}
\end{equation}
where $\Phi$ is the dilaton, $\chi$ stands for the axion. By
setting $\chi=-e^{-\Phi}+\chi_0$, the dilaton term and the axion
term can cancel. Then the solution of (\ref{action}) can be
written as \cite{KG1}
\begin{equation}
ds^2=e^{\frac{\Phi}{2}}[-\frac{r^2}{R^2}f(r)dt^2+\frac{r^2}{R^2}d\vec{x}^2+\frac{1}{f(r)}\frac{R^2}{r^2}dr^2],\label{metric}
\end{equation}
with
\begin{equation}
e^\Phi=1+\frac{q}{r^4_t}log\frac{1}{f(r)},\qquad
f(r)=1-\frac{r_t^4}{r^4}\label{efai},
\end{equation}
where $R$ is the AdS radius with $R^4=4\pi
g_sN_c{\alpha^\prime}^2$. $\vec{x}=x_1,x_2,x_3$ are the boundary
coordinates. $r$ stands for the radial coordinate. The event
horizon and the boundary are located at $r=r_t$ and $r=\infty$,
respectively. The Hawking temperature is $T=r_t/{(\pi R^2)}$. The
parameter $q$ refers to the D-instanton density as well as the
vacuum expectation value of the gauge field condensate. Note that
for $q=0$ in (\ref{metric}) the $AdS_5$-Schwarzschild metric is
reproduced.

The next step is to make the pair moving. Supposing that the
plasma is at rest and the frame is moving in one direction, i.e.,
the $x_3$ direction, so that
\begin{equation}
dt=dt^\prime cosh\beta-dx_3^\prime sinh\beta, \qquad
dx_3=-dt^\prime sinh\beta+dx_3^\prime cosh\beta,\label{tr}
\end{equation}
where $\beta$ is called the velocity or rapidity.

Inserting (\ref{tr}) into (\ref{metric}) and dropping the primes,
one obtains the boosted metric as
\begin{eqnarray}
ds^2
=e^{\frac{\Phi}{2}}[(-\frac{r^2}{R^2}f(r)cosh^2\beta+\frac{r^2}{R^2}sinh^2\beta)dt^2-2sinh\beta
cosh\beta(\frac{r^2}{R^2}-\frac{r^2}{R^2}f(r))dtdx_3\nonumber\\+(-\frac{r^2}{R^2}f(r)sinh^2\beta+\frac{r^2}{R^2}cosh^2\beta)dx_3^2+\frac{r^2}{R^2}(dx_1^2+dx_2^2)+\frac{R^2}{r^2}f(r)^{-1}dr^2].\label{metric1}
\end{eqnarray}

Noticed that for $\beta=0$ in (\ref{metric1}) the metric of
(\ref{metric}) is recovered.

\section{imaginary potential}
We now follow the argument in \cite{JN} to investigate the
imaginary potential of a moving quark anti-quark pair for the
background metric (\ref{metric1}). In general, to analyze the
moving case, one needs to consider different alignments for the
pair with respect to the plasma wind, i.e., transverse
$(\theta=\pi/2)$, parallel $(\theta=0)$, or arbitrary direction
$(\theta)$. In this work, we discuss two extreme cases:
$\theta=\pi/2$ and $\theta=0$.

\subsection{Transverse to the wind $(\theta=\pi/2)$}
In this subsection, we consider the system perpendicularly to the
wind in the $x_1$ direction, the coordinates are parameterized by
\begin{equation}
t=\tau, \qquad x_1=\sigma,\qquad x_2=0,\qquad x_3=0,\qquad
r=r(\sigma),\label{par}
\end{equation}
where the quark and anti-quark are located at $x_1=-\frac{L}{2}$
and $x_1=\frac{L}{2}$ with $L$ the inter-distance.

The string action is
\begin{equation}
S=-\frac{1}{2\pi\alpha^\prime}\int d\tau d\sigma\mathcal
L=-\frac{1}{2\pi\alpha^\prime}\int d\tau d\sigma\sqrt{-g},
\label{S}
\end{equation}
where $g$ is the determinant of the induced metric with
\begin{equation}
g_{\alpha\beta}=g_{\mu\nu}\frac{\partial
X^\mu}{\partial\sigma^\alpha} \frac{\partial
X^\nu}{\partial\sigma^\beta},
\end{equation}
with $g_{\mu\nu}$ the metric and $X^\mu$ the target space
coordinates.

Substituting (\ref{par}) into (\ref{metric1}), one obtains the
induced metric as
\begin{equation} g_{00}=e^{\frac{\Phi}{2}}[\frac{r^2}{R^2}f(r)cosh^2\beta-\frac{r^2}{R^2}sinh^2\beta], \qquad
g_{11}=e^{\frac{\Phi}{2}}[\frac{r^2}{R^2}+\frac{R^2}{f(r)r^2}\dot{r}^2],
\end{equation}
with $\dot{r}=dr/d\sigma$.

Then the corresponding lagrangian density can be written as
\begin{equation}
\mathcal L=\sqrt{a(r)+b(r)\dot{r}^2},
\end{equation}
with
\begin{eqnarray}
&a(r)&=e^{\Phi}[\frac{r^4}{R^4}f(r)cosh^2\beta-\frac{r^4}{R^4}sinh^2\beta],\nonumber\\&b(r)&=e^{\Phi}[cosh^2\beta-\frac{1}{f(r)}sinh^2\beta].
\end{eqnarray}

Note that the action does not depend on $\sigma$ explicitly, so
the Hamiltonian is a constant,
\begin{equation}
\mathcal L-\frac{\partial\mathcal
L}{\partial\dot{r}}\dot{r}=constant.
\end{equation}

Considering the boundary condition at $\sigma=0$,
\begin{equation}
\dot{r}=0,\qquad r=r_c,
\end{equation}
one finds
\begin{equation}
\dot{r}=\frac{dr}{d\sigma}=\sqrt{\frac{a^2(r)-a(r)a(r_c)}{a(r_c)b(r)}}\label{dotr},
\end{equation}
with
\begin{equation}
a(r_c)=e^{\Phi(r_c)}[\frac{r_c^4}{R^4}f(r_c)cosh^2\beta-\frac{r_c^4}{R^4}sinh^2\beta],
\end{equation}
\begin{equation}
f(r_c)=1-\frac{r_t^4}{r_c^4}, \qquad
e^{\Phi(r_c)}=1+\frac{q}{r^4_t}log\frac{1}{f(r_c)},
\end{equation}
where $r=r_c$ is the deepest point of the U-shaped string.

Integrating (\ref{dotr}), the inter-distance of $Q\bar{Q}$ reads
\begin{equation}
L=2\int_{r_c}^{\infty}dr\sqrt{\frac{a(r_c)b(r)}{a^2(r)-a(r)a(r_c)}}\label{x}.
\end{equation}

Substituting (\ref{dotr}) into (\ref{S}) and subtracting the self
energy of the two quarks \cite{JMM,ABR,SJR}, the real part of the
heavy quark potential is obtained as
\begin{equation}
ReV_{Q\bar{Q}}=\frac{1}{\pi\alpha^\prime}\int_{r_c}^{\infty}dr[\sqrt{\frac{a(r)b(r)}{a(r)-a(r_c)}}-\sqrt{b(r_0)}]-
\frac{1}{\pi\alpha^\prime}\int_{r_t}^{r_c}dr\sqrt{b(r_0)}\label{re},
\end{equation}
with $b(r_0)=b(r\rightarrow\infty)$.

On the other hand, using the thermal worldsheet fluctuation method
\cite{JN}, the imaginary part of the potential is found to be
\begin{equation}
ImV_{Q\bar{Q}}=-\frac{1}{2\sqrt{2}\alpha^\prime}[\frac{a^\prime(r_c)}{2a^{\prime\prime}(r_c)}-\frac{a(r_c)}{a^\prime(r_c)}]\sqrt{b(r_c)},\label{im}
\end{equation}
with
\begin{eqnarray}
a^\prime(r_c)=e^{\Phi^\prime(r_c)}r_c^4[f(r_c)cosh^2\beta-sinh^2\beta]+e^{\Phi(r_c)}[4r_c^3f(r_c)cosh^2\beta+r_c^4f'(r_c)cosh^2\beta-4r_c^3sinh^2\beta]\label{1},
\end{eqnarray}
\begin{eqnarray}
a^{\prime\prime}(r_c)&=&e^{\Phi^{\prime\prime}(r_c)}r_c^4[f(r_c)cosh^2\beta-sinh^2\beta]+e^{\Phi^\prime(r_c)}[8r_c^3f(r_c)cosh^2\beta-8r_c^3sinh^2\beta+2r_c^4f^\prime(r_c)cosh^2\beta]\nonumber\\
&+&e^{\Phi(r_c)}[12r_c^2f(r_c)cosh^2\beta+8r_c^3f^\prime(r_c)cosh^2\beta+r_c^4f^{\prime\prime}(r_c)cosh^2\beta-12r_c^2sinh^2\beta]\label{2},
\end{eqnarray}
\begin{equation}
b(r_c)=e^{\Phi(r_c)}[cosh^2\beta-\frac{1}{f(r_c)}sinh^2\beta],
\end{equation}
\begin{eqnarray}
e^{\Phi^\prime(r_c)}=-qr_t^{-4}\frac{f^\prime(r_c)}{f(r_c)log10},\qquad
e^{\Phi^{\prime\prime}(r_c)}=-qr_t^{-4}\frac{f^{\prime\prime}(r_c)f(r_c)-f^\prime(r_c)f^\prime(r_c)}{f^2(r_c)log10},\label{3}
\end{eqnarray}
\begin{eqnarray}
f^\prime(r_c)=4r_t^4r_c^{-5},\qquad
f^{\prime\prime}(r_c)=-20r_t^4r_c^{-6},\qquad
\end{eqnarray}
where, for simplicity, we have taken $R=1$. One can check that for
$q=0$ in (\ref{im}) the result of \cite{MAL} is recovered. And for
$q=\beta=0$ in (\ref{im}), the formula of \cite{JN} is reproduced
as well.

Before evaluating the imaginary potential, we should pause here to
discuss the regime of applicability of this model. It was argued
\cite{JN,JN2} that there are three restrictions to the formula
(\ref{im}). First, the term $b(r_c)$ needs to be positive, which
leads to
\begin{equation}
\varepsilon<\varepsilon_{max,1}=(1-tanh^2\beta)^{1/4},
\end{equation}
with $\varepsilon\equiv r_t/r_c$.

Second, the imaginary potential should be negative, so
\begin{equation}
\frac{a^\prime(r_c)}{2a^{\prime\prime}(r_c)}-\frac{a(r_c)}{a^\prime(r_c)}>0,
\label{c2}
\end{equation}
which yields
\begin{equation}
\varepsilon>\varepsilon_{min},
\end{equation}
here the length expression of $\varepsilon_{min}$ is not shown,
but one will see its behavior in fig.2 below.

The third restriction is related to the maximum value of the
inter-distance. To address this, we plot $LT$ versus $\varepsilon$
for various cases in fig.1, one can see that in each plot there
exists a maximum value of $LT_{max}$, which corresponds to a
critical value of $\varepsilon_{max,2}$, and that $LT$ is a
increasing function of $\varepsilon$ for
$\varepsilon<\varepsilon_{max,2}$ but a decreasing one for
$\varepsilon>\varepsilon_{max,2}$. Actually, for the later case it
is necessary to consider highly curved configurations which are
not solutions of the Nambu-Goto action \cite{DB}, so here we are
mostly interested in the region of
$\varepsilon<\varepsilon_{max,2}$. Also, it is interesting to
mention that $LT_{max}$ has been used to define the dissociation
length for the moving $Q\bar{Q}$, see e.g. \cite{HL,HL1}. But in
this work, we tend to use it to define the region of applicability
of the U-shaped string configuration, as follows from \cite{JN2}.

By combining the three restrictions, we plot $\varepsilon_{min},
\varepsilon_{max,1}, \varepsilon_{max,2}$ versus $\beta$ for
$q=0.2$ in fig.2. Other cases with different values of $q$ have
similar picture. From the figures, one can see that these
restrictions lead to a narrow window. Or in other words, the
regime of applicability of the formula (\ref{im}) is
$\varepsilon_{min}<\varepsilon<\varepsilon_{max,2}$.

\begin{figure}
\centering
\includegraphics[width=8cm]{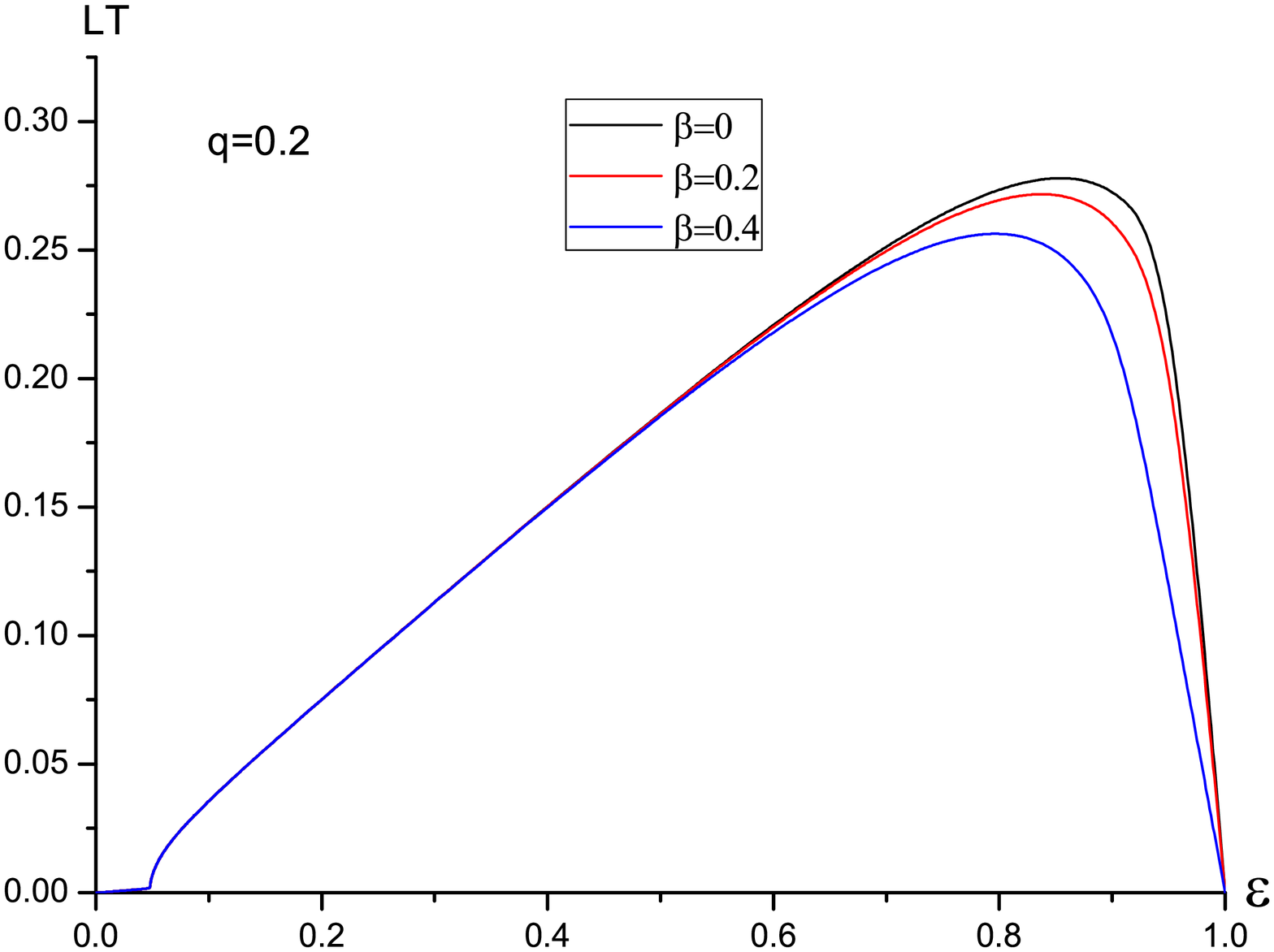}
\includegraphics[width=8cm]{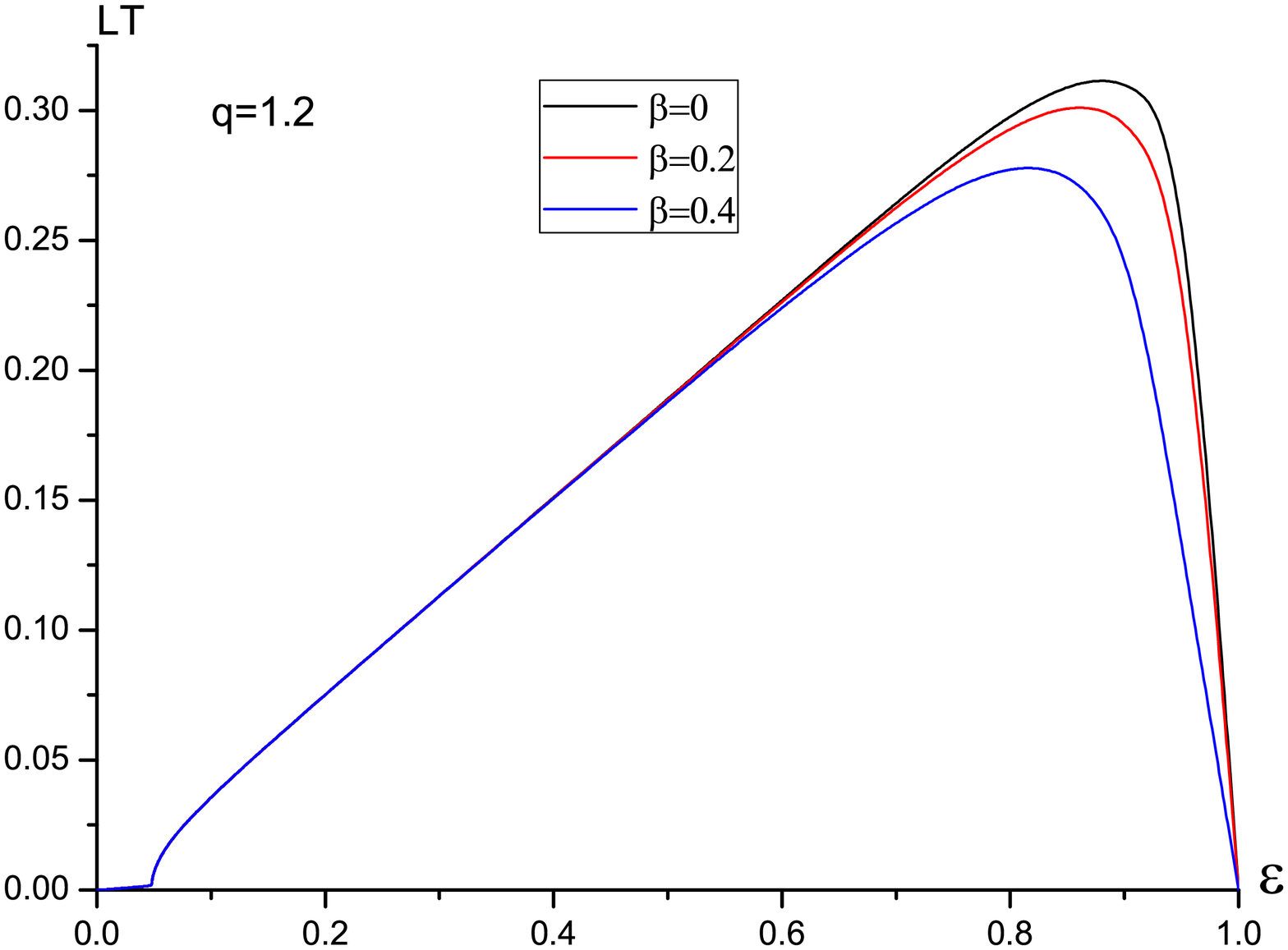}
\caption{Plots of $LT$ versus $\varepsilon$ for $\theta=\pi/2$.
Left: $q=0.2$. Right: $q=1.2$. In all of the plots from top to
bottom $\beta=0,0.2,0.4$, respectively.}
\end{figure}

\begin{figure}
\centering
\includegraphics[width=10cm]{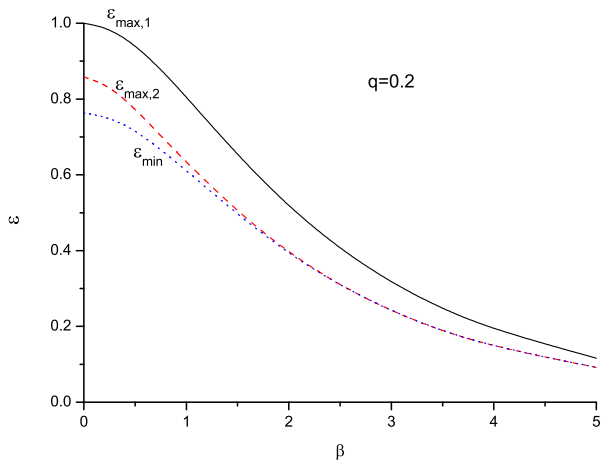}
\caption{Limiting curves of $\varepsilon$ versus $\beta$ for
$\theta=\pi/2$ and $q=0.2$. From top to bottom, the three curves
represent $\varepsilon_{max,1}$ (solid line),
$\varepsilon_{max,2}$ (dash line) and $\varepsilon_{min}$ (dot
line), respectively.}
\end{figure}

To proceed, we study the effects of the velocity and the D-density
on the inter-distance. From fig.1, one sees that at a fixed $q$,
increasing $\beta$ the value of $LT_{max}$ decreases. Namely, the
velocity has the effect of decreasing the inter-distance,
consistently with the findings of \cite{JN2,MAL}. Moreover,
comparing the left panel with the right one, one finds that at a
fixed $\beta$, increasing $q$ leads to increasing the value of
$LT_{max}$. Therefore, one concludes that the D-density affects
the inter-distance in the opposite way of the velocity.

Next, we investigate the imaginary potential in the region of
$\varepsilon_{min}<\varepsilon<\varepsilon_{max,2}$. In Fig.3, we
plot ImV$_{Q\bar{Q}}/(\sqrt{\lambda}T)$ versus $LT$ for two
different values of $q$. The left panel is plotted for $q=0.2$
while the right one is for $q=1.2$. In both panels from left to
right $\beta=0.4,0.2,0$, respectively. From the figures, one can
see that the imaginary potential starts at a $L_{min}$,
corresponding to $\varepsilon=\varepsilon_{min}$ or
Im$V_{Q\bar{Q}}=0$, and ends at a $L_{max}$, corresponding to
$\varepsilon_{max,2}$. Also, one finds that at a fixed $\beta$,
increasing $q$ the imaginary potential decreases. As discussed in
\cite{JN2}, the dissociation properties of quarkonia should be
sensitive to the imaginary potential, and if the onset of
ImV$_{Q\bar{Q}}$ happens for smaller $LT$, the suppression will be
stronger. Thus, we conclude that the presence of the D-instanton
density makes the suppression weaker, reverse to the effect of the
velocity. Interestingly, it was shown \cite{ZQ1} that the
D-instanton density has the effect of suppressing the heavy quark
potential thus making the quarkonia melt harder, in agreement with
the findings here.

\begin{figure}
\centering
\includegraphics[width=8cm]{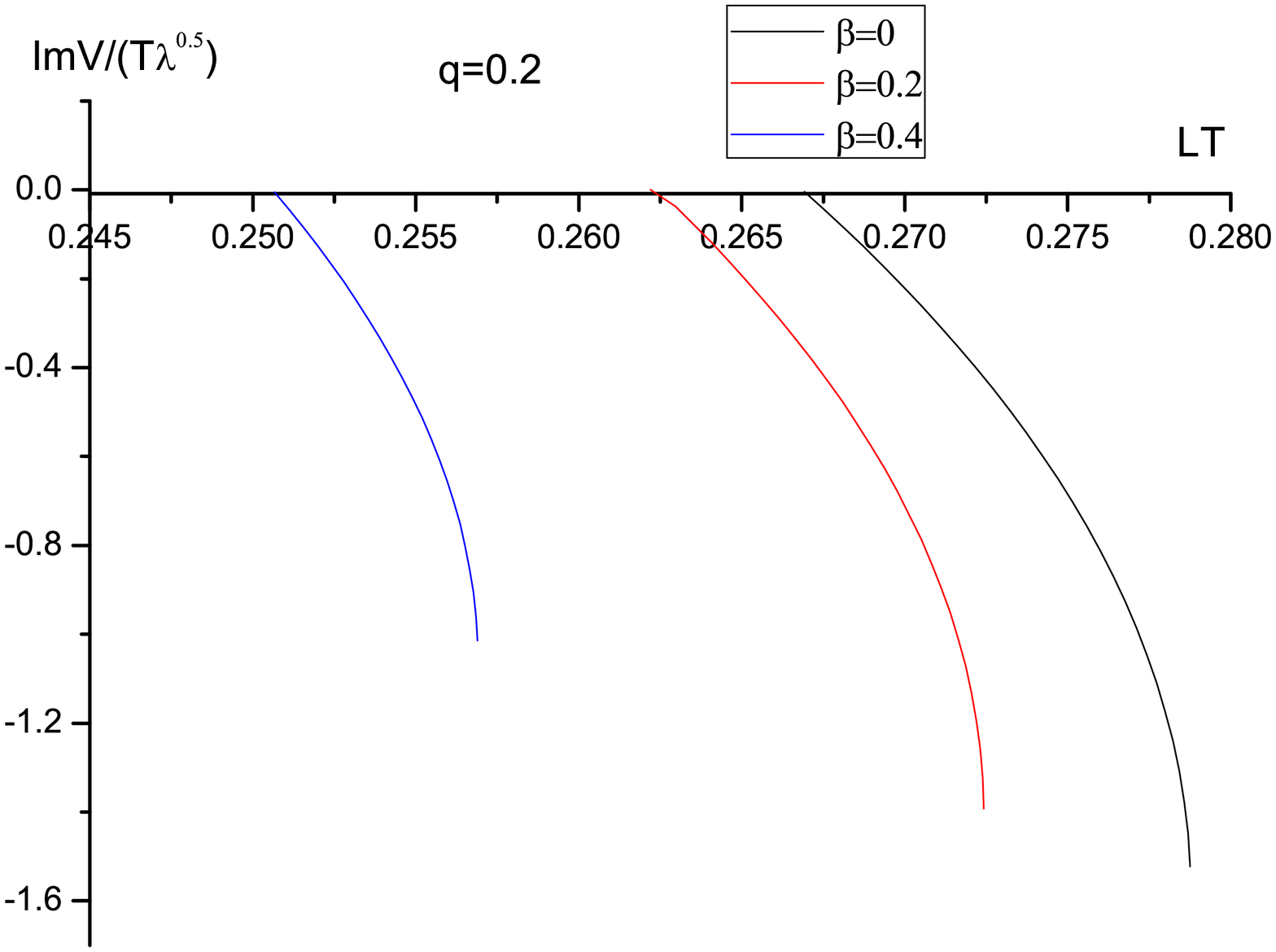}
\includegraphics[width=8cm]{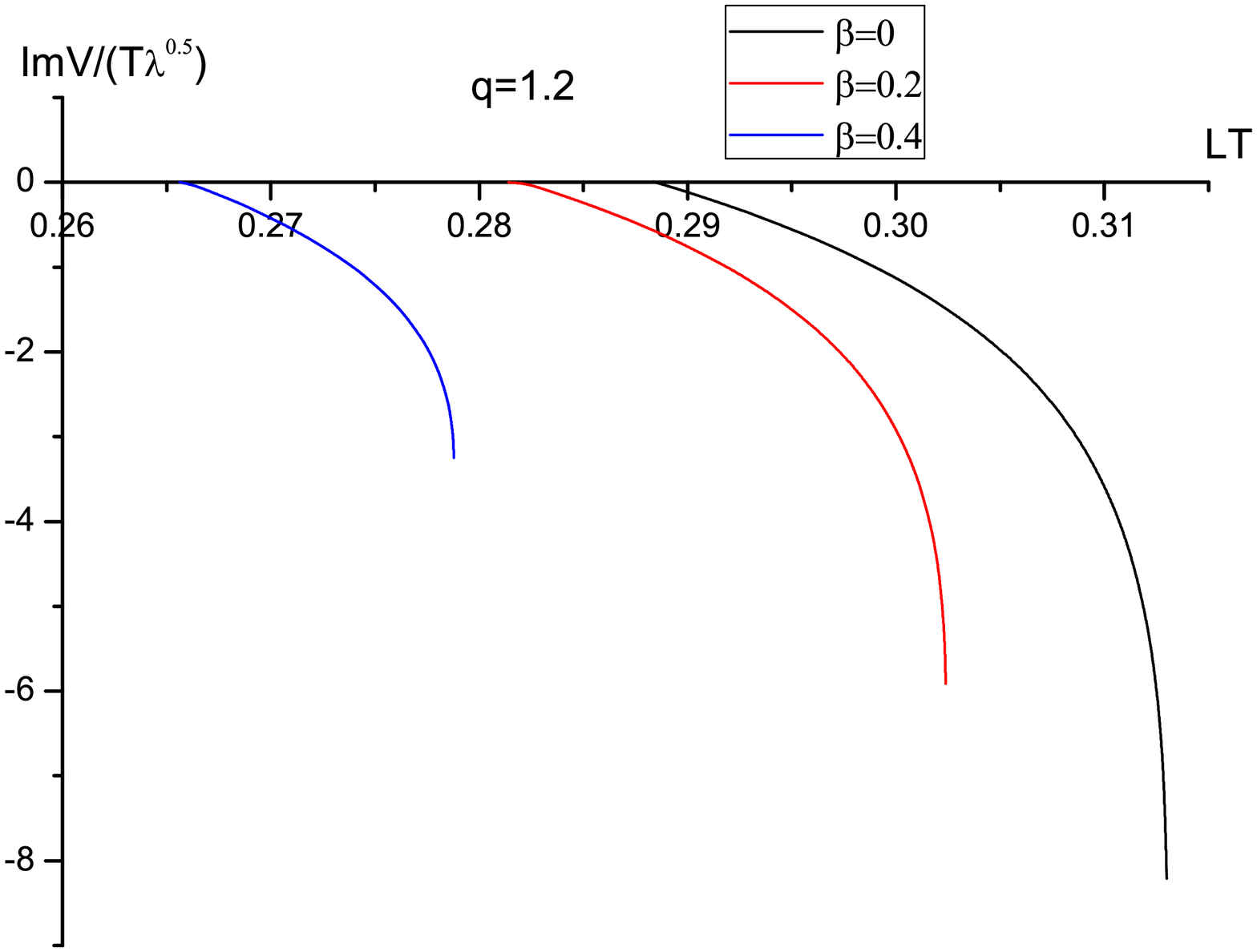}
\caption{Plots of $ImV/(\sqrt{\lambda}T)$ versus $LT$ for
$\theta=\pi/2$. Left: $q=0.2$. Right: $q=1.2$. In all of the plots
from left to right $\beta=0.4,0.2,0$, respectively.}
\end{figure}

\subsection{Parallel to the wind $(\theta=0)$}
In this subsection we discuss the system parallel to the wind. The
coordinates are parameterized as
\begin{equation}
t=\tau, \qquad x_1=0,\qquad x_2=0,\qquad x_3=\sigma,\qquad
r=r(\sigma),\label{par1}
\end{equation}
where the quark and anti-quark are located at $x_3=-\frac{L}{2}$
and $x_3=\frac{L}{2}$, respectively.

The next analysis is similar to the previous subsection, so we
just present the final results. The inter-distance is
\begin{equation}
L=2\int_{r_c}^{\infty}dr\sqrt{\frac{A(r_c)B(r)}{A^2(r)-A(r)A(r_c)}}\label{x1},
\end{equation}
with
\begin{equation}
A(r)=e^{\Phi}\frac{r^4}{R^4}[f(r)sinh^4\beta+f(r)cosh^4\beta-sinh^2\beta
cosh^2\beta(1+f^2(r))],
\end{equation}
\begin{equation}
A(r_c)=e^{\Phi(r_c)}\frac{r_c^4}{R^4}[f(r_c)sinh^4\beta+f(r_c)cosh^4\beta-sinh^2\beta
cosh^2\beta(1+f^2(r_c))],
\end{equation}
\begin{equation}
B(r)=e^{\Phi}[cosh^2\beta-\frac{1}{f(r)}sinh^2\beta].
\end{equation}

The imaginary potential is
\begin{equation}
ImV_{Q\bar{Q}}=-\frac{1}{2\sqrt{2}\alpha^\prime}[\frac{A^\prime(r_c)}{2A^{\prime\prime}(r_c)}-\frac{A(r_c)}{A^\prime(r_c)}]\sqrt{B(r_c)},\label{im1}
\end{equation}
with
\begin{eqnarray}
A^\prime(r_c)&=&e^{\Phi(r_c)}[(sinh^4\beta+cosh^4\beta)(4r_c^3f(r_c)+r_c^4f^\prime(r_c))-sinh^2\beta
cosh^2\beta(4r_c^3+4r_c^3f^3(r_c)\nonumber\\&+&2r_c^4f(r_c)f^\prime(r_c))]+e^{\Phi^\prime(r_c)}r_c^4[f(r_c)sinh^4\beta+f(r_c)cosh^4\beta-sinh^2\beta
cosh^2\beta(1+f^2(r_c))]\label{11},
\end{eqnarray}
\begin{eqnarray}
A^{\prime\prime}(r_c)&=&e^{\Phi^{\prime\prime}(r_c)}r_c^4[f(r_c)sinh^4\beta+f(r_c)cosh^4\beta-sinh^2\beta
cosh^2\beta(1+f^2(r_c))]\nonumber\\&+&2e^{\Phi^\prime(r_c)}[(sinh^4\beta+cosh^4\beta)(4r_c^3f(r_c)+r_c^4f^\prime(r_c))-sinh^2\beta
cosh^2\beta(4r_c^3+4r_c^3f^3(r_c)+2r_c^4f(r_c)f^\prime(r_c))]\nonumber\\&+&e^{\Phi}(r_c)[(sinh^4\beta+cosh^4\beta)(12r_c^2f(r_c)+8r_c^3f^\prime(r_c)+r_c^4f^{\prime\prime}(r_c))-sinh^2\beta
cosh^2\beta(12r_c^2+12r_c^2f^2(r_c)\nonumber\\&+&8r_c^3f(r_c)f^\prime(r_c)+2r_c^4f(r_c)f^{\prime\prime}(r_c)+8r_c^3f^\prime(r_c)f(r_c)+2r_c^4f^\prime(r_c)f^\prime(r_c))],
\end{eqnarray}
\begin{equation}
B(r_c)=e^{\Phi(r_c)}[cosh^2\beta-\frac{1}{f(r_c)}sinh^2\beta],
\end{equation}
and
\begin{eqnarray}
e^{\Phi^\prime(r_c)}=-qr_t^{-4}\frac{f^\prime(r_c)}{f(r_c)log10},\qquad
e^{\Phi^{\prime\prime}(r_c)}=-qr_t^{-4}\frac{f^{\prime\prime}(r_c)f(r_c)-f^\prime(r_c)f^\prime(r_c)}{f^2(r_c)log10},\label{31}
\end{eqnarray}
\begin{eqnarray}
f^\prime(r_c)=4r_t^4r_c^{-5},\qquad
f^{\prime\prime}(r_c)=-20r_t^4r_c^{-6}.\qquad
\end{eqnarray}

Likewise, we plot $LT$ versus $\varepsilon$ and
ImV$_{Q\bar{Q}}/(\sqrt{\lambda}T)$ versus $LT$ for $\theta=0$ in
fig.4 and fig.5, respectively. From these figures, one finds that
the results are very similar to the transverse case: the
D-instanton density increases the inter-distance and decreases the
imaginary potential, while the velocity has opposite effects.

Moreover, to compare the effects of the D-instanton density on the
imaginary potential between $\theta=0$ and $\theta=\pi/2$, we plot
ImV$_{Q\bar{Q}}/(\sqrt{\lambda}T)$ versus $LT$ with $\beta=0.4$
for two different $q$ in the left panel of fig.6. The left two
adjacent curves correspond to
ImV$_{Q\bar{Q},\perp}/(\sqrt{\lambda}T)$ (lower line) and
ImV$_{Q\bar{Q},\parallel}/(\sqrt{\lambda}T)$ (upper line) for
$q=0.2$ while the right ones correspond to their counterparts for
$q=1.2$. One sees that at fixed $\beta$, increasing $q$, the
differences between the onsets of the
ImV$_{Q\bar{Q},\perp}/(\sqrt{\lambda}T)$ and
ImV$_{Q\bar{Q},\parallel}/(\sqrt{\lambda}T)$ become more and more
bigger. This behavior can also be seen from the right panel of
fig.6, which shows $L_{min}$ versus $q$. One finds that as $q$
increases, $L_{min}$ increases, which indicates that the
suppression becomes weaker. On the other hand, the results show
that as $q$ increases, the differences between $L_{min,\perp}$ and
$L_{min,\parallel}$ widen, especially at large $\beta$. Also, at
fixed $\beta$, the slope of $L_{min,\perp}$ is smaller than that
of $L_{min,\parallel}$, which implies that the D-instanton density
has smaller effects for the transverse case rather than parallel,
differs from the velocity, which has stronger effects for the
perpendicular case \cite{MAL,KB1}.

\begin{figure}
\centering
\includegraphics[width=8cm]{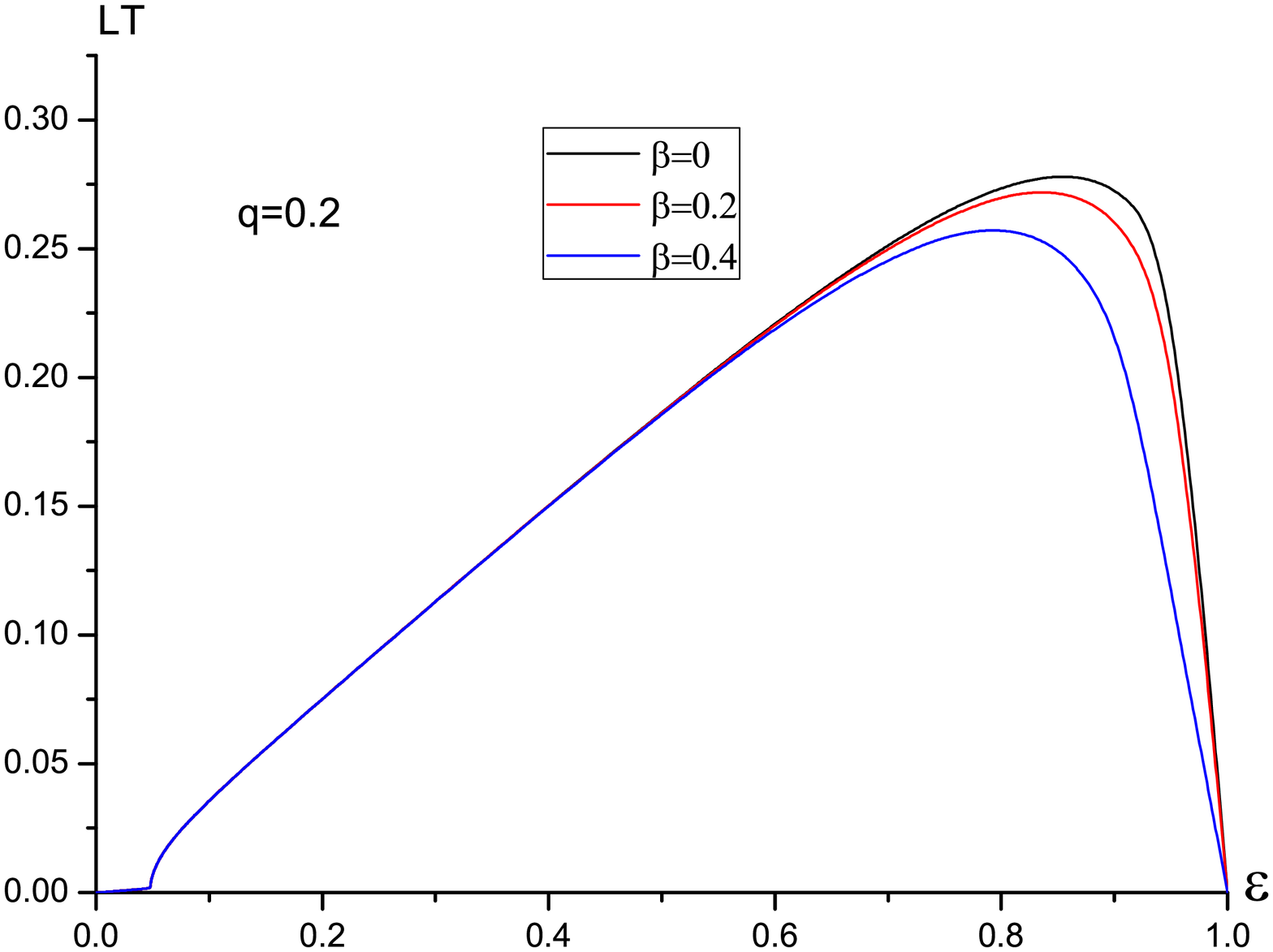}
\includegraphics[width=8cm]{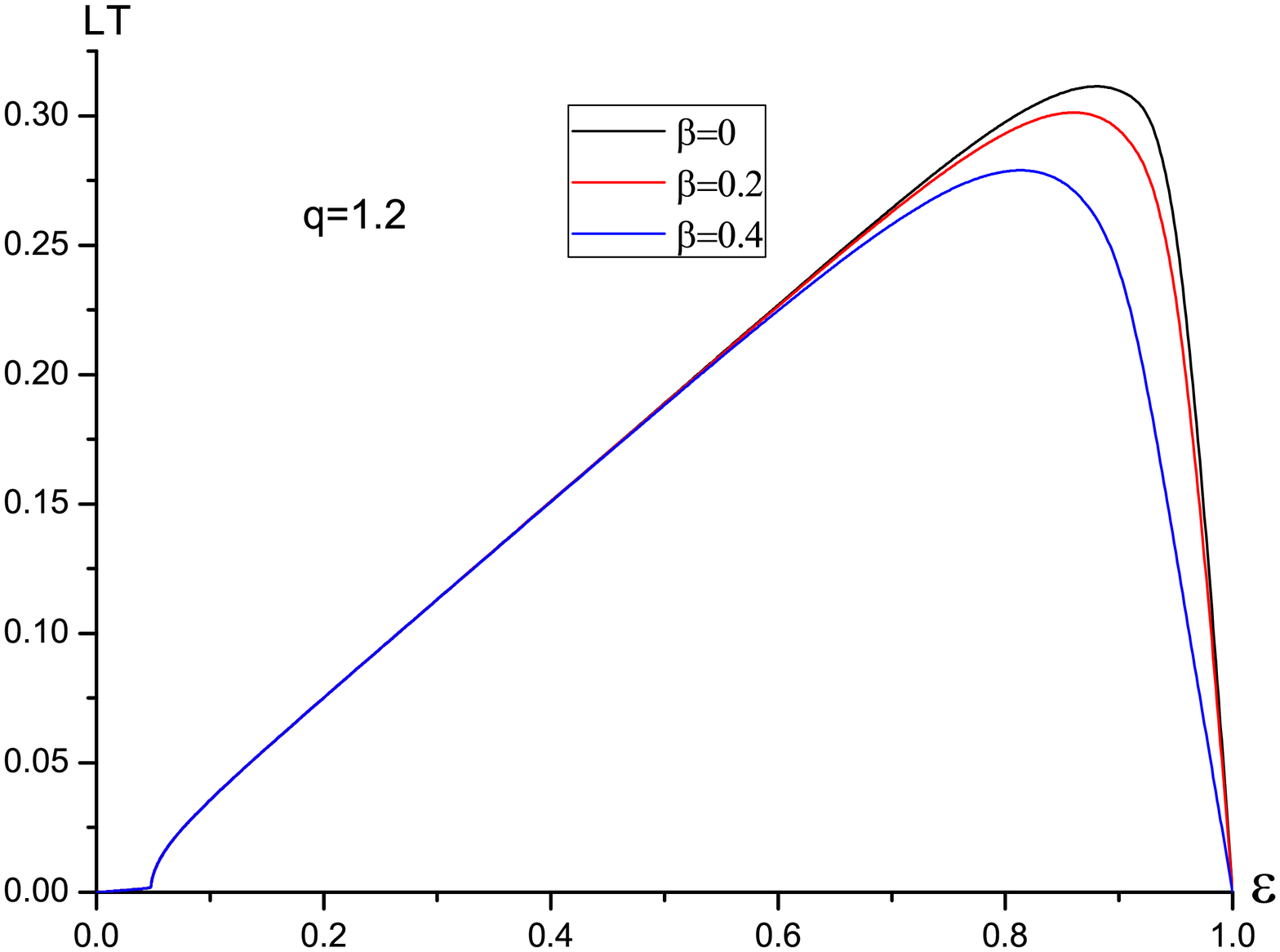}
\caption{Plots of $LT$ versus $\varepsilon$ for $\theta=0$. Left:
$q=0.2$. Right: $q=1.2$. In all of the plots from top to bottom
$\beta=0,0.2,0.4$, respectively.}
\end{figure}

\begin{figure}
\centering
\includegraphics[width=8cm]{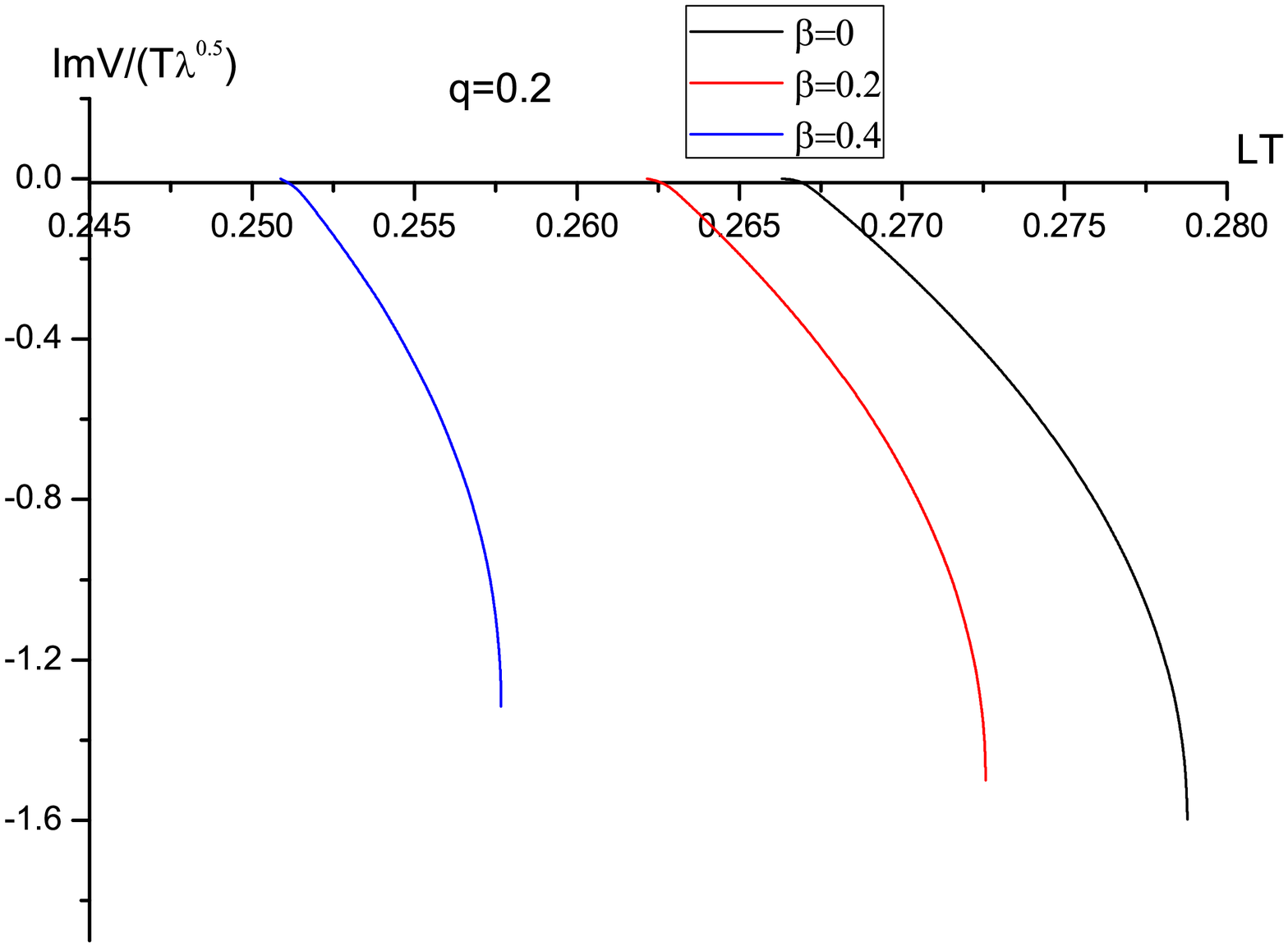}
\includegraphics[width=8cm]{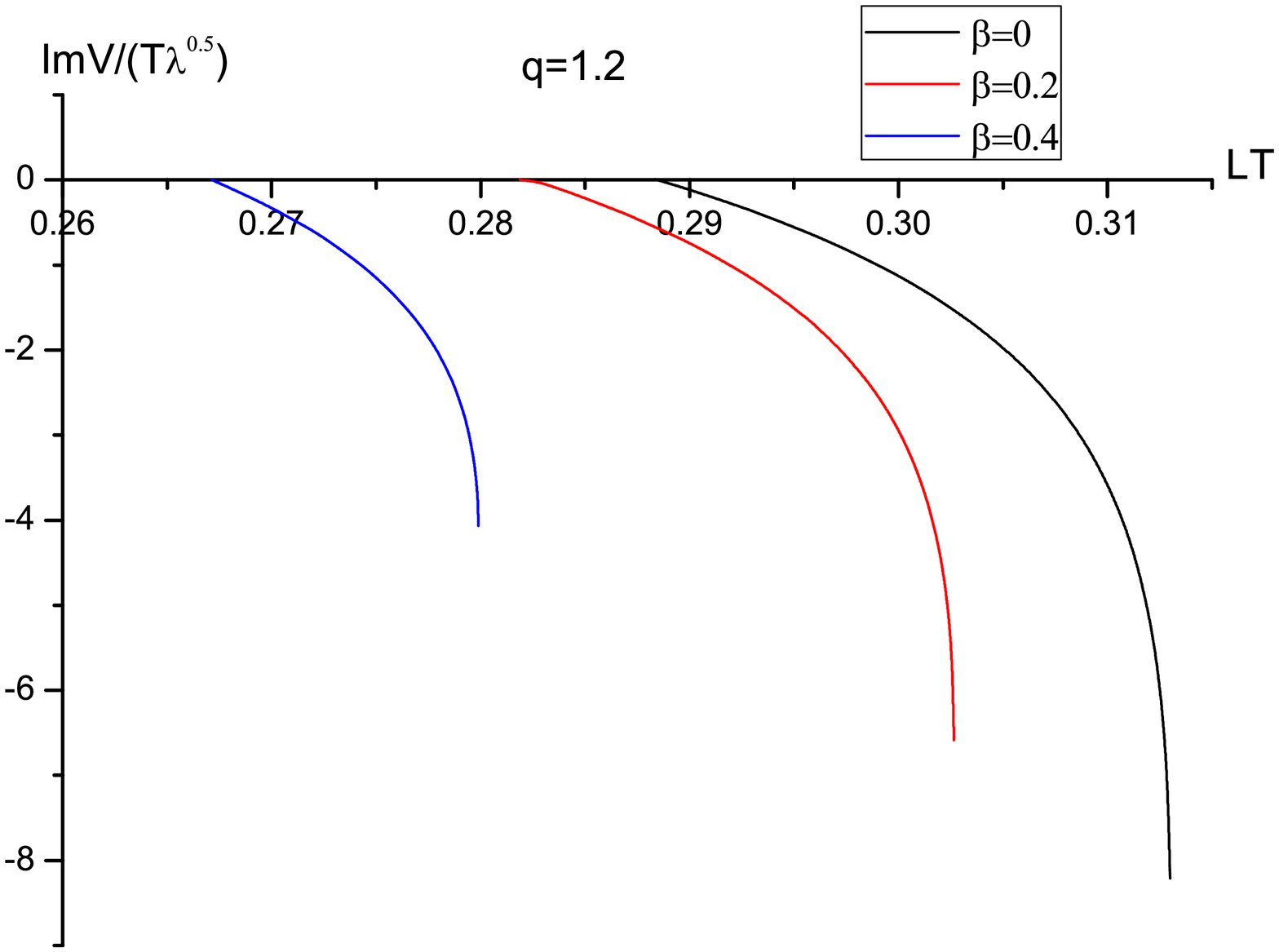}
\caption{Plots of $ImV/(\sqrt{\lambda}T)$ versus $LT$ for
$\theta=0$. Left: $q=0.2$. Right: $q=1.2$. In all of the plots
from left to right $\beta=0.4,0.2,0$, respectively.}
\end{figure}

\begin{figure}
\centering
\includegraphics[width=8cm]{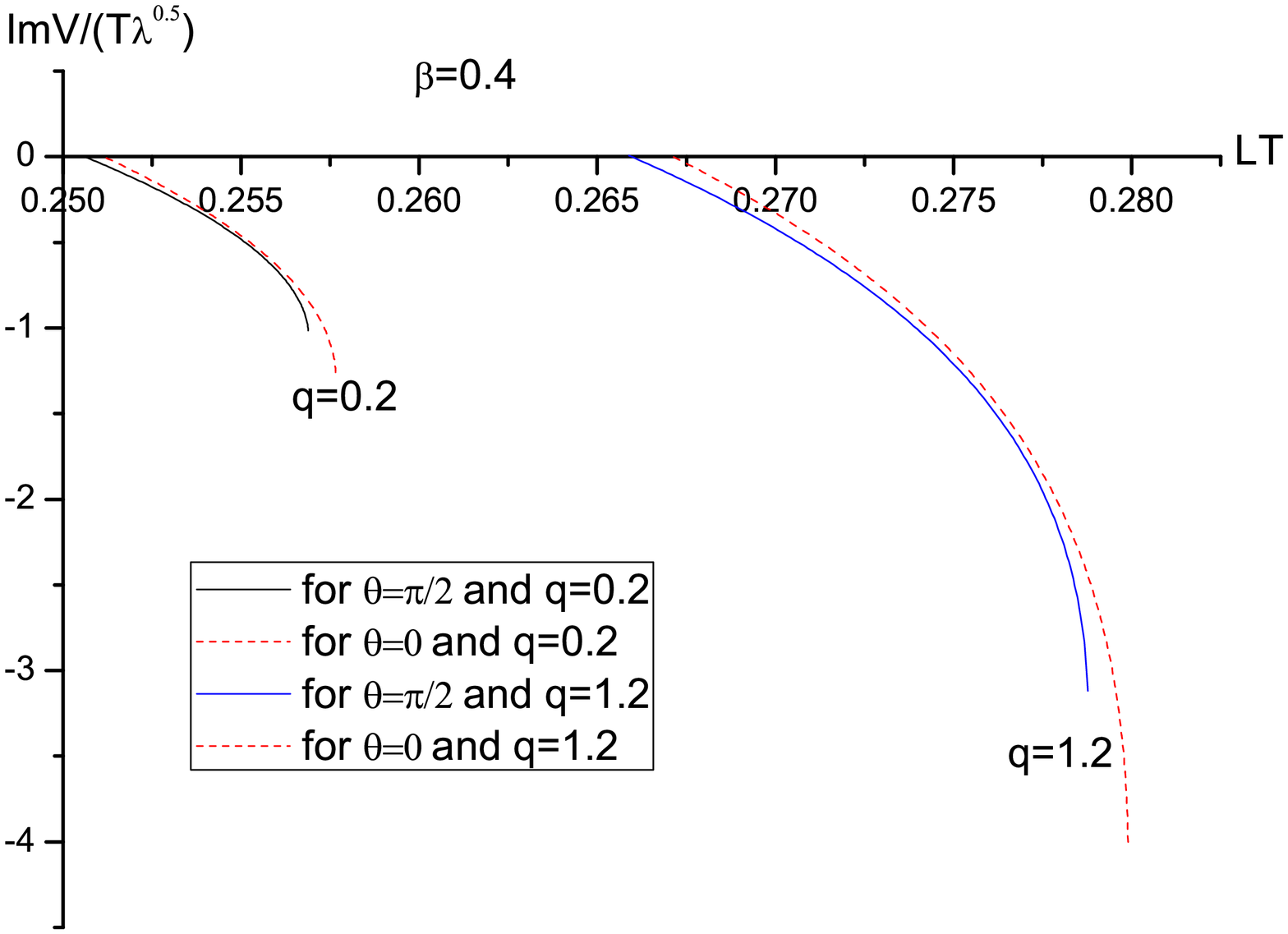}
\includegraphics[width=9cm]{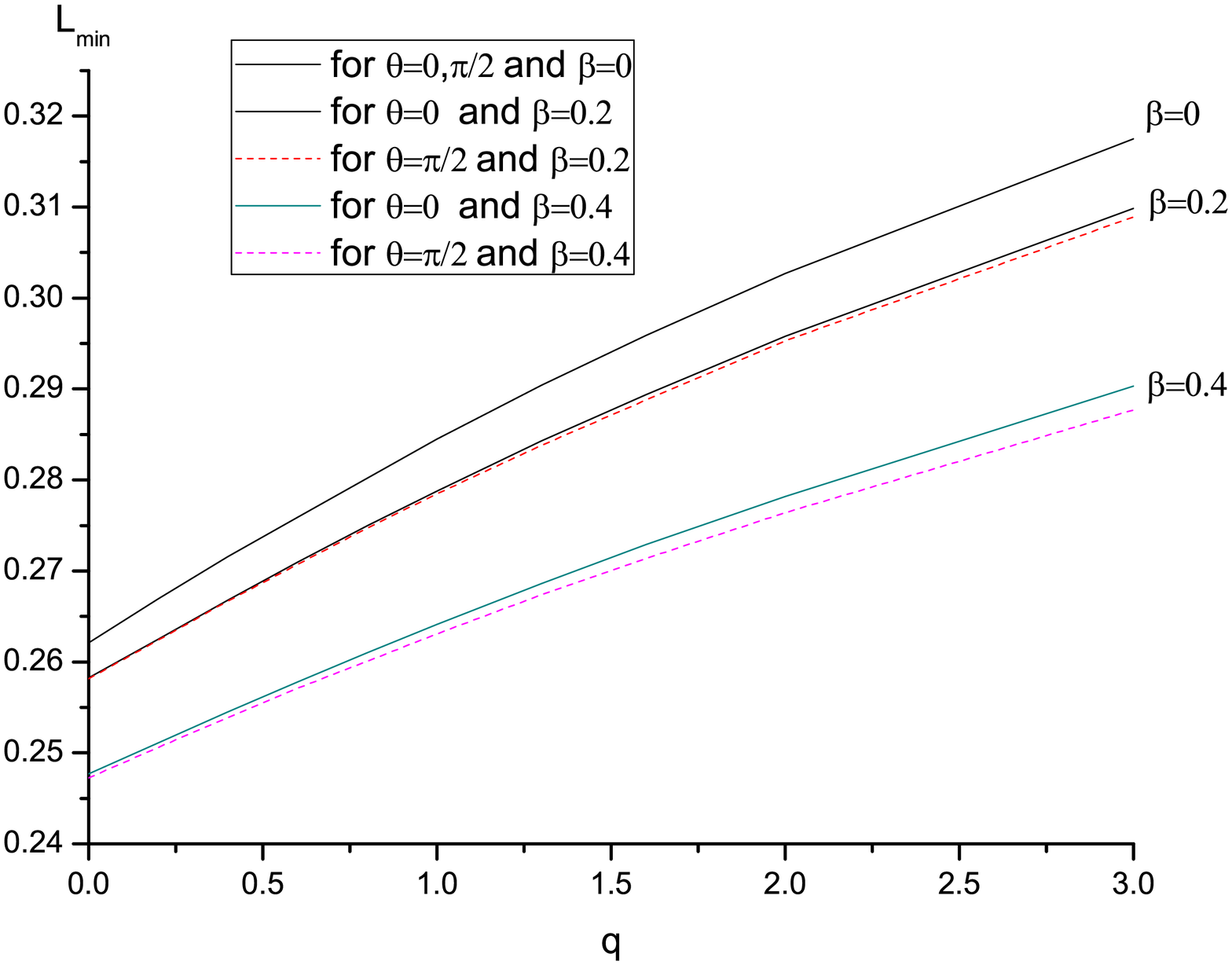}
\caption{Left: $ImV/(\sqrt{\lambda}T)$ versus $LT$ for
$\beta=0.4$. Right: $LT_{min}$ versus $q$ for $\beta=0,0.2,0.4$,
respectively.}
\end{figure}

\section{conclusion and discussion}
In this paper, we studied the imaginary potential of moving
quarkonia in a D-instanton background, generated by a dilaton
field, corresponding to D-instanton density contributions. The
features of the background configuration are similar to QCD at
finite temperature. Thus, one expects the results obtained from
this theory should shed qualitative insights into analogous
questions in QCD.

To analyze the imaginary potential, we considered the quark
anti-quark pair moving transverse and parallel to the plasma wind,
respectively. For both cases, it is found that the D-instanton
density increases the inter-distance and decreases the imaginary
potential, reverse to the effect of the pair's velocity. Since the
dissociation properties of quarkonia should be sensitive to the
imaginary potential, and the larger the ImV$_{QQ}$, the stronger
the suppression. Thus, one concludes that the presence of the
D-instanton density makes the suppression weaker, reverse to the
effect of the velocity. Furthermore, it is shown that the
D-instanton density has stronger effects for the parallel case
rather than perpendicular.

Finally, it would be interesting to mention that the entropic
force has been recently argued to be responsible for melting heavy
quarkonia \cite{DEK,KHA}. It is also of interest to study this
quantity in the D-instanton background. We leave this for further
study.

\section{Acknowledgments}

This research is partly supported by the Ministry of Science and
Technology of China (MSTC) under the ¡°973¡± Project No.
2015CB856904(4). Z-q Zhang is supported by NSFC under Grant No.
11705166. G.C is supported by NSFC under Grant No. 11475149. D-f.
Hou is partly supported by the NSFC under Grants Nos. 11735007,
11375070, 11521064.


\end{document}